\documentclass[11pt]{article}

\usepackage[a4paper,margin=1in]{geometry}

\usepackage[english]{babel}
\usepackage[utf8]{inputenc}
\usepackage[T1]{fontenc}
\usepackage{csquotes}
\usepackage[hyphens]{url}
\usepackage{hyperref}

\usepackage{multicol}
\usepackage{caption}

\usepackage{chngpage}

\usepackage{amsmath}
\usepackage{mathtools}
\usepackage{graphicx}
\usepackage{xspace}
\usepackage{booktabs}
\usepackage{pifont} 
\usepackage{microtype} 
\usepackage{soul}

\usepackage{authblk} 

\newcommand{\footnotewithoutmarker}[1]{{
  \let\thempfn\relax
  \footnotetext[0]{#1}
}}

\usepackage[
	style=numeric-comp,
	sorting=none,
	hyperref,
	singletitle=true,
	backend=biber]{biblatex}

\newcommand{\Screentocamera}{QRCode-flashing\xspace}
\newcommand{\screentocamera}{QRCode-flashing\xspace}

\newcommand{\nbParticipants}{$364$\xspace}
\newcommand{\proP}{$64\%$\xspace}
\newcommand{\weeklyNeedP}{$61\%$\xspace}

\newcommand{\notMonthlyNeedP}{$12\%$\xspace}
\newcommand{\usageBTP}{$6\%$\xspace}
\newcommand{\satBTP}{$17\%$\xspace}
\newcommand{\usageCloudP}{$74\%$\xspace}
\newcommand{\satCloudP}{$75\%$\xspace}
\newcommand{\usageExtP}{$34\%$\xspace}
\newcommand{\satExtP}{$67\%$\xspace}
\newcommand{\usageAirdropP}{$42\%$\xspace}
\newcommand{\satAirdropP}{$60\%$\xspace}

\title{A critical review of mobile device-to-device communication}

\author[1]{Lauric Desauw}
\author[1]{Adrien Luxey-Bitri}
\author[1]{Rémy Raes}
\author[1]{Romain Rouvoy}
\author[2]{Olivier Ruas}
\author[1]{Walter Rudametkin}

\affil[1]{Univ.\,Lille, Inria, CNRS, UMR\,9189\,CRIStAL, France}
\affil[2]{Pathway, \texttt{pathway.com}, France}

\newcommand{\myfootertext}{
    \footnotesize
    Published in August 2023.\\
    Authors listed in alphabetical order. Adrien Luxey-Bitri is the corresponding author.\\
    E-mail adresses: 
    Lauric Desauw: \texttt{lauric.desauw@tutanota.com}; 
    Adrien Luxey-Bitri: \texttt{adrien.luxey@inria.fr}; 
    Rémy Raes: \texttt{remy.raes@inria.fr}; 
    Romain Rouvoy: \texttt{romain.rouvoy@inria.fr}; 
    Olivier Ruas: \texttt{olivier@pathway.com}; 
    Walter Rudametkin: \texttt{walter.rudametkin@inria.fr}.
}

\begin{document}

\maketitle
\footnotewithoutmarker{\myfootertext}

\begin{abstract}
Since the advent of mobile devices, both end-users and the IT industry have been longing for direct \emph{device-to-device} (D2D) communication capabilities, expecting new kinds of interactive, personalized, and collaborative services.
Fifteen years later, many D2D solutions have been implemented and deployed, but their availability and functionality are underwhelming.
Arguably, the most widely-adopted D2D use case covers the pairing of accessories with smartphones;
however, many other use cases---such as mobile media sharing---did not progress.
Pervasive computing and cyber-physical convergence need local communication paradigms to scale.
For inherently local use cases, they are even more appealing than ever:
eschewing third-parties simultaneously fosters environmental sustainability, privacy and network resiliency.
This paper proposes a survey on D2D communication, investigates its deployment and adoption, with the objective of easing the creation and adoption of modern D2D frameworks.
We present the results of an online poll that estimates end-users' utilisation of D2D processes, and review enabling technologies and security models.
\end{abstract}

\section{Introduction}
\emph{Device-to-device} (D2D) communication refers to the capability of mobile devices to communicate directly with one another, eschewing intermediaries, such as a cellular base station.
The D2D paradigm has been physically enabled for a long time by a wide range of wireless communication channels (e.g. infrared, Wi-Fi, Bluetooth, NFC, UWB), 
but it failed to be largely adopted by user-facing applications. 
It is currently hindered by technical (and non-technical) barriers that make it an unsuitable choice for developers and end-users alike.
Most networked applications thus rely on centralized infrastructures, even for inherently local use cases, 
such as proximity sensing and face-to-face media exchanges---which limits the capacity of pervasive cyber-physical systems to scale~\cite{cookPervasiveComputingScale2012,contiInternetPeopleIoP2017}.
D2D communication would have an important part to play in modern communication, however:
due to its low infrastructural environmental footprint, inherent privacy, 
and for the pathway it paves to pervasive services beyond the Internet.
This article thus proposes a critical survey of D2D's state of the art, focused on gathering insights to further its appropriation by the community.
\textit{What is the current deployment and adoption status of the device-to-device stack, what are its capabilities, and how can it be leveraged on widely available commodity hardware?}
To answer these questions, we will report on:
\textit{(i)} the results of an online poll on D2D usage conducted by our team, featuring \nbParticipants participants;
\textit{(ii)} a review of the D2D communication channels that are the most widely available to end-users;
\textit{(iii)} an analysis of the D2D's paradigm security model.

\begin{figure}[t]
	\centering
	\includegraphics[width=0.4\columnwidth]{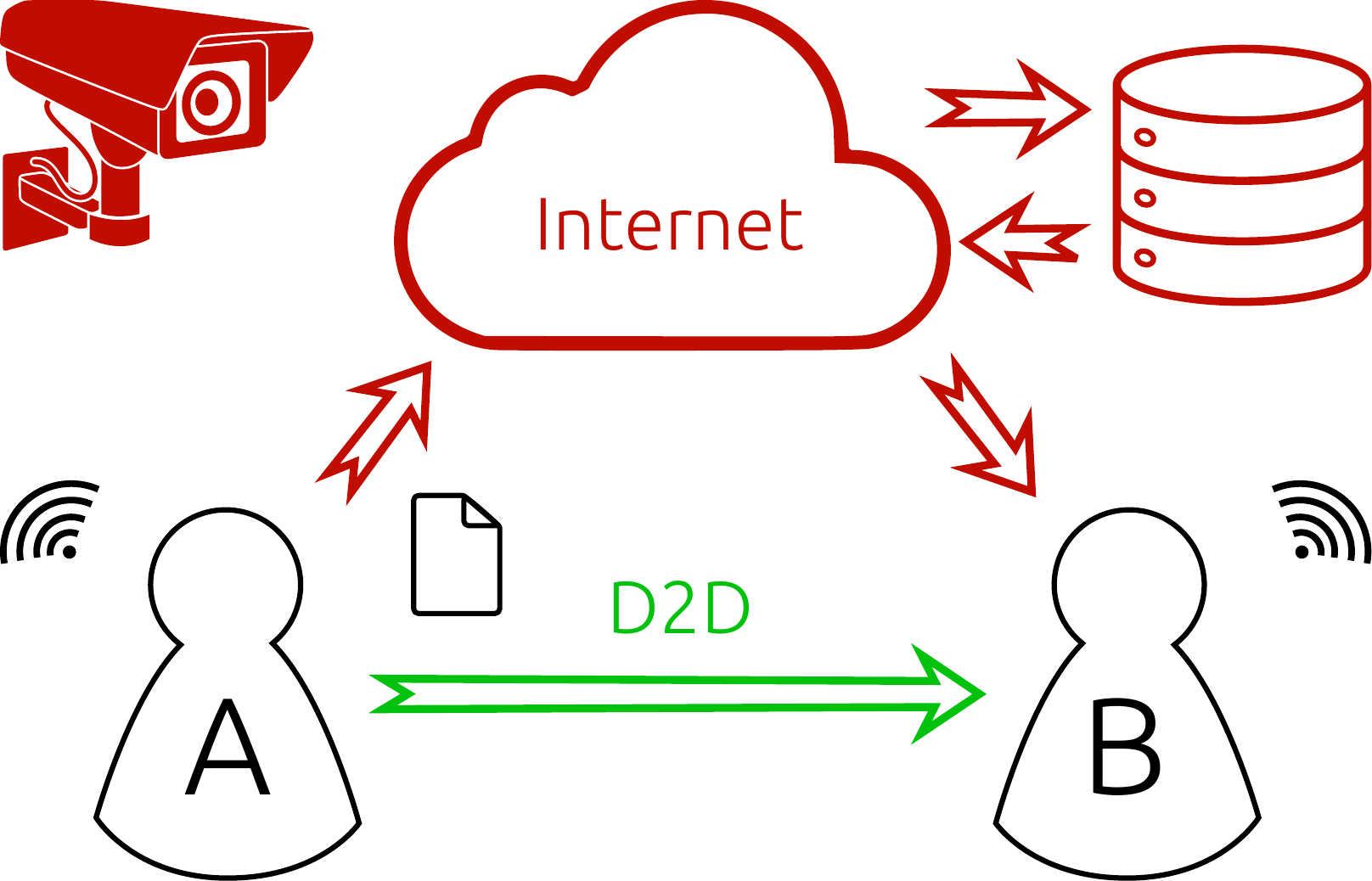}
	\caption{For an inherently D2D use case such as transferring a file between nearby users, resorting to an Internet service for the exchange leads to more energy consumption, more storage space used, and exposes the users to digital surveillance (degraded privacy).}
	\label{fig:d2d_vs_cloud}
\end{figure}

\subsection{Scope}
`Device-to-device communication' is an umbrella term that has many definitions.
In this article, we frame a definition of D2D focused on user-centric use cases.
We consider a D2D communication as an exchange of information that happens \emph{wirelessly}, in a \emph{single hop} (without intermediaries), using channels and protocols already available on commodity \emph{end-user mobile devices}.
Intermediaries, such as a cellular base station, might assist in the establishment of the connection, but should not participate beyond that.
This article does not cover \emph{machine-to-machine} (M2M) applications, although some M2M-oriented protocols will be presented. 
Indeed, automatic communication between machines is not user-centric, and generally does not leverage the same protocols as the ones available on end-user equipment.

\subsection{D2D scenarios}

Jameel~\emph{et~al.}~\cite{jameel_survey_2018} proposed a taxonomy of D2D use cases, namely traffic offloading, providing emergency services, extending cellular coverage, reliable health monitoring, mobile tracking and positioning, and data dissemination.
Among these use cases, we do not address provisioning emergency services, as it leverages multi-hop communication through advanced routing. 
Health monitoring and mobile tracking are left aside, because they utilize specialized sensor networks outside the end-users' realm (e.g. M2M communication).
Traffic offloading and extension of cellular coverage revolve around facilitating access to Internet services, and are thus not our focus---although they will be mentioned.
Finally, our paper specifically targets \textit{user-centric data dissemination}, from direct file sharing to collaborative social services, such as crowdsourced environment sensing.
An example of the file-sharing scenario is given in figure~\ref{fig:d2d_vs_cloud} where, in contrast to the use of Internet services, two users who share a file directly achieve better energy efficiency, less resource consumption and better privacy (cf. sec.~\ref{sec:motivation}).
Our online survey, presented in sec.~\ref{sec:d2d_survey}, shows that participants face this use case on a regular basis.

This rest of this paper is organised as follows: 
Section~\ref{sec:motivation} first motivates the need for D2D communication.
In Section~\ref{sec:d2d_survey}, we report on an online poll that was conducted by our team on D2D usage, and analyse its results.
Section~\ref{sec:d2d_channels} then describes the physical channels enabling D2D communication, while Section~\ref{sec:security_models} presents the D2D security model and its properties.
Section~\ref{sec:conclusion} finally discusses our findings and concludes this paper.

\section{Motivation}\label{sec:motivation}
Networked user applications are thriving, and Internet traffic is growing at an astonishing rate.\footnotemark~
Yet, nearly all users' network sessions transit through the \emph{cloud}: a cohort of data centers that feature high-grade equipments and are physically connected as close as possible to the Internet's `backbone'.
\footnotetext{See Cisco's Annual Internet Reports~\cite{cisco_annual_report_2020} and Wikipedia's `Internet Traffic' page for aggregated data~\cite{internet_traffic_wikipedia}.}
In this section, we develop several arguments that advocate for more D2D communications in mobile device networking.

\subsection{Supporting the energy transitions} 
\label{ssec:environment}
We first consider the example depicted in figure~\ref{fig:d2d_vs_cloud}, where Alice and Bob, who are physically close, want to share a file. 
Unless both users own Apple products that feature AirDrop, as of now, the typical `netizen' will adopt an indirect solution by uploading files to an online service (e.g. file storage, e-mail, instant communication) before their recipient can fetch them. 
The alternative is more direct and leverages a D2D protocol (e.g.~Bluetooth, Wi-Fi Direct), such that the file is immediately transmitted from Alice's smartphone to Bob's, without the intervention of any intermediaries.
How is the indirect solution more costly?
To answer, we need to break down a online-based file exchange:
Alice's file needs to be transmitted from her smartphone to her network gateway (either her Wi-Fi access point or her cellular base station), and further transmitted over the Internet to the data centre where the online service will store the file on a persistent storage.
Then, Bob needs to connect to the same service, so the file performs approximately the same journey back from the data centre's storage, over the Internet to Bob's network gateway, and to Bob's smartphone.
This \emph{sharing-as-a-service} process supported by the cloud is costly in several regards:
\begin{itemize}
	\item \emph{Data in transit}: On its way to \& from the data centre, the file hops through an unbounded amount of network equipments (routers, switches, cables).
	Although the electrical consumption and related carbon footprint of data in transfer are hard to measure and individually negligible, this typical use case---globally---induces a non-negligible load on the network~\cite{priceGoogleDriveNow2017}, leading to the necessity to keep increasing and maintaining the Internet's throughput;

	\item \emph{Network congestion}: Increasing the Internet's throughput imposes to build more infrastructures, producing more routers and cables, consuming more electricity, and consequently extracting more raw materials from the Earth~\cite{international_energy_agency_iea_role_2021};

	\item \emph{Cold storage}: The third-party server has to store a copy of the file that could remain there for an unknown time~\cite{cheng_over_2012}.
	The service provider thus requires an ever-increasing storage capacity (e.g., hard drives) to sustain their customers' demands.
\end{itemize}

By shifting direct communication between user devices, the aforementioned costs would be highly limited. 
Only the incompressible energy and storage required to locally transfer the file between the two devices would be consumed. 
In essence, to lower the environmental impact of the Internet, it is crucial to shorten communication chains, wherever possible.
We postulate that---given the digitalization of our lives and the variety of use cases that could eschew a cloud third-party---the raw material savings enabled by D2D communication are substantial, thanks to its lower electrical consumption and negligible use of infrastructure.

\subsection{Unplugging the surveillance economy}

Every time Alice makes a connection or \emph{request} to a data center to use an online service, the following actors obtain details of the communication:
\begin{itemize}
	\item The network operators (notably Alice's \emph{Internet Service Provider} (ISP) or cellular network provider), as well as the data centre's owner, obtain Alice's IP address, the potentially encrypted payload, its size, and the request timestamp. 
	If the connection is not encrypted, they can also read the request's content.
	Even when encryption is enforced, a more intrusive operator may also be able to obtain information about Alice's device or to inspect the content of her requests---for example through \emph{Deep Packet Inspection} (DPI)~\cite{asghari_deep_2013};

	\item In addition to the above, the cloud service operator obtains application-specific information, as well as the payload of the request. 
	Following our file sharing use case, the cloud service obtains the filename, size, content, and likely the recipient's address or identifier (e.g. their e-mail address).
\end{itemize}

Network service providers are legally bound to retain this data for security purposes~\cite{la_quadrature_du_net_resume_2020}.
Besides this legal requirement, Shoshana Zuboff~\cite{zuboff_age_2019} describes how the digital economy revolves around the accumulation and processing of personal data for financial benefit. 
Although each datum is seemingly harmless, this `surveillance economy' poses serious threats to individuals~\cite{al-saggaf_use_2015}, free speech~\cite{nakashima_for_2013}, and democratic societies at large~\cite{cadwalladr_i_2018,matz_psychological_2017,luxey_e-squads_2019_political_risks}.
By eschewing third parties, D2D communication holds great potential to prevent the accumulation and marketization of personal information and its detrimental effects.

\subsection{Uneven access to a broadband Internet connection}
\label{ssec:bad_connectivity}
Worldwide, access to the Internet and to mobile devices is highly uneven:
in 2021 almost half of the World's population is kept offline, according to the United Nations~\cite{UN_half_offline}.
Surprisingly, $94\%$ of people are within coverage of a mobile broadband network~\cite{GSM_coverage}.
In addition to the half billion people without coverage, billions more who could have access cannot afford it.
The geographical distribution of people without access to the Internet shows that Africa is the continent of most concern~\cite{internet_stats}, but there are local disparities within many countries~\cite{vogels2021some}.

People's budgets may be the main cause of this digital divide.
Indeed, users need to buy a device and, arguably more importantly, they must pay service fees to access the Internet.
Prices vary widely per country and depend on many factors.
The impact on consumers also depends on their purchasing power.
And, while unlimited data plans do exist, most people still pay per gigabyte consumed or have limited plans.
In 2022, comparing China to the United States shows a $6.4$ times difference in pricing, with average prices of \$0.41 and \$5.62~USD per GB of mobile data, respectively~\cite{internet_prices}.
However, countries with very similar culture and purchasing power can also have very different data rates, as shown through France and Belgium, two neighbouring European countries.
In 2021, France had a median price per GB of \$0.41~USD, while in Belgium the median price was \$5.28~USD/GB: $12.8$ times more expensive.
As data usage increases due to the growing volumes of content (such as video streaming), connection fees can be a burden to many, including in developed countries, such as the U.S.~\cite{internet_price_covid}.
By enabling proximity networking without fees, D2D communication would be a way to decrease those costs and to include the discarded of the digital divide.
Local communications can also provide basic connectivity in situations where Internet coverage is unavailable or inconsistent (e.g. in planes, inside buildings or in disaster situations).

\section{The discouraging D2D user experience}\label{sec:d2d_survey}
\newcommand{\nbComments}{82\xspace}

We conducted an online survey on `how people usually share documents between physically close devices', available at 
\url{https://d2dsurvey.net}.
along with the survey answers' raw data.
This use case could typically be undertaken using D2D protocols.
Our goal was to estimate how the public achieves it, and their satisfaction using the different solutions available.
After a short inquiry of their IT know-how, participants were asked about the frequency with which they need to exchange documents between nearby devices.
Then, they were asked to express how often they use, and their level of satisfaction with regard to Bluetooth, external storage (ext), cloud services, and lastly AirDrop for Apple users only.
A free-text field allowed participants to provide comments and feedback.
We gathered \nbParticipants answers, including \proP professionals of the IT community.

\begin{figure}[t] 
\begin{minipage}{\textwidth} 
\begin{multicols}{2}
  \centering
  \includegraphics[scale=0.6]{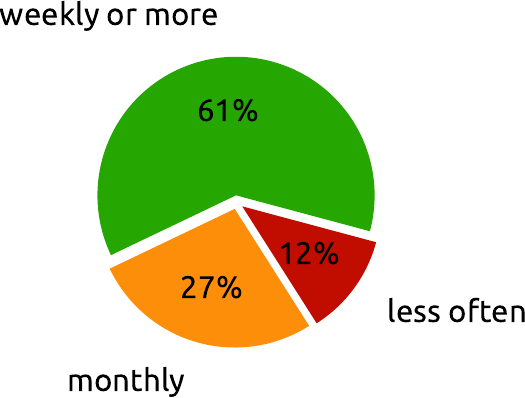}
  
  \raggedright
  \captionof{figure}{Distribution of the frequency of the need for D2D communication. \weeklyNeedP of the surveyed users need D2D communication at least once a week, while only \notMonthlyNeedP need it less than once a month.}
  \label{fig:survey:needD2D}
  \vspace{2ex}

  \centering
  \includegraphics[width=0.6\columnwidth]{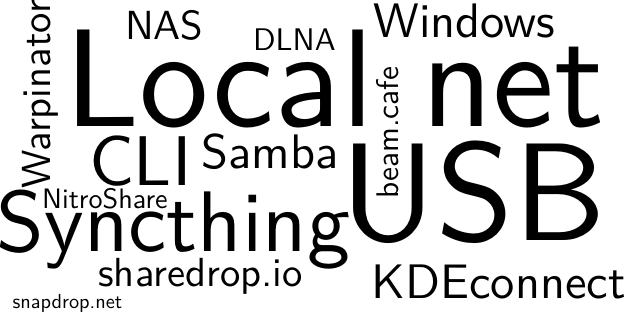}
  
  \raggedright
  \captionof{figure}{Alternate solutions used by participants of our survey for close file sharing, 
  proposed in the free-text field.
  Technologies are not exclusive, e.g. \emph{Samba} is generally a \emph{Windows} \emph{Local net} solution.
  Font-size is proportional to the proposition's frequency.}
  \label{fig:survey:alternatives}

\columnbreak

  \centering
  \includegraphics[scale=0.7]{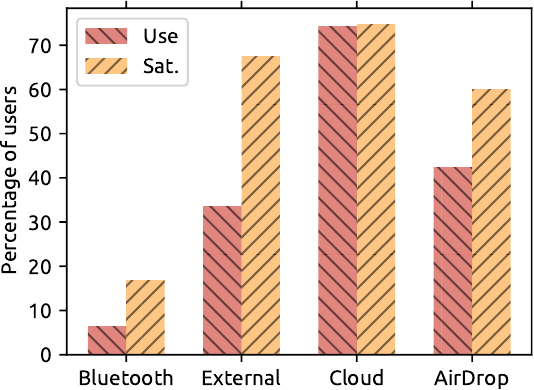}
  \raggedright
  \captionof{figure}{Use and satisfaction (Sat.) of the different available technologies.
  For the use, only users frequently using the given technology are taken into account.
  Bluetooth is the least popular technology to exchange documents between nearby devices, despite being designed for this purpose.
  On the other hand, the cloud is the preferred solution, while it is the only not relying on the device proximity.}
  \label{fig:survey:comparatif}

\end{multicols}
\end{minipage}
\end{figure}

\paragraph{There is a need for D2D communication.}
Figure~\ref{fig:survey:needD2D} shows the distribution of the frequency of the need to share files between nearby devices.
\weeklyNeedP of the participants face the use case weekly or more, while only \notMonthlyNeedP need D2D file sharing less than once a month.
This figure highlights the ubiquity of media exchange in our digital lifestyles: from lolcats and memes, to text documents (administrative, press...), including voice recordings, souvenir pictures and videos, etc.

\paragraph{Many alternate solutions to a simple problem}
Out of the \nbParticipants participants, \nbComments left an answer in the free-text feedback field. 
As shown in figure~\ref{fig:survey:alternatives}, people solve the close file sharing issue within their entourage using many alternate solutions: D2D or not, ad-hoc or purposeful.
Notably: Syncthing\footnotemark, Samba, \texttt{\href{https://www.sharedrop.io/}{sharedrop.io}}, KDE Connect, Warpinator, proximity sharing with a Wi-Fi Hotspot (Local net), \texttt{rsync} (CLI), etc.
This variety can be interpreted in different ways.
On the one hand, it shows the versatility of the digital tool, as it enables many different ways to solve a simple task.
On the other hand, it also highlights the lack of a universal D2D solution for file exchange.

\footnotetext{ 
Syncthing exists precisely to allow file exchange without involving third-parties.
It works over the Internet, or in a D2D fashion if the involved devices share a local network.
See \url{https://syncthing.net/}.
}

\paragraph{Existing D2D technologies are not satisfactory.}
Figure~\ref{fig:survey:comparatif} reports on the percentage of participants who frequently use each technology and their level of satisfaction.
Except for AirDrop, the satisfaction percentages are computed on the total number of participants and not only on those frequently using the given technology, since not being satisfied about a technology has an impact on its usage frequency.
For AirDrop, the percentages are computed only for Apple users.

Only \usageBTP use Bluetooth frequently, and only \satBTP are satisfied with it.
While AirDrop seems to be a better D2D alternative for Mac users, \usageAirdropP of Mac users use it frequently and \satAirdropP deem it satisfactory.
Non-D2D technologies are significantly preferred.
\usageExtP of the participants frequently use external storage, with \satExtP satisfaction.
The most frequently used technology to exchange documents between nearby devices is, ironically, the cloud: \usageCloudP of the participants frequently use it for that purpose, and \satCloudP are satisfied with it.

A first observation is that, despite the need for D2D file sharing, people rarely use Bluetooth or Airdrop. 
They prefer other sharing means (external storage, or mostly the cloud).
Airdrop has a usage and a satisfaction percentage of \usageAirdropP and \satAirdropP respectively, among Apple-using participants.
Both statistics fall below that of the cloud.
The lack of inter-compatibility with other vendors may partly explain this underwhelming observation.
However, the real question lies in the minor \satBTP satisfaction towards Bluetooth, and its even lower \usageBTP usage rate.
This disuse is surprising, given that this channel is notably built for the very purpose of exchanging files.
Regarding external storage, its \usageExtP usage rate can be explained by the burden of using a storage proxy, including cables. 
However, participants are fairly satisfied with it: once plugged in, cabled solutions are indeed reliable.
Finally, regarding the cloud:
its \satCloudP satisfaction rate is in line with the amount of investment and the many actors involved in network-based file exchange.
However, following our paper's motivation, we deem that \emph{\usageCloudP of the participants regularly resorting to the cloud for such a D2D scenario is bad news}.
For the sake of the environmental footprint of ICT, privacy and users' data bills, 
professionals ought to work towards reducing this proportion in favour of D2D channels.
The rest of this article presents communication channels and security models to this end.

\section{Communication channels}\label{sec:d2d_channels}
The \emph{radio frequency} (RF) spectrum is regulated in every country: it is forbidden to emit on specific frequency bands without authorisation.
Consequently, each RF channel resides on an agreed-upon frequency band.
In the first subsection, we introduce the D2D channels existing in radio bands that are free to use without authorisation, called \emph{unlicensed} or \emph{out-band} frequency bands.
In subsection~\ref{ssec:inband_channels}, we turn our attention to D2D communication channels that make use of \emph{licensed} cellular networks, hence requiring a subscription.
Subsection~\ref{ssec:proprietary} then presents several proprietary solutions that leverage upon physical channels in order to provide user-friendly D2D connectivity.
Finally, subsection~\ref{ssec:hardware} explains how all these channels are made available to the users on the hardware level.

Table~\ref{tab:channels} features a compilation of communication channels that were considered the most readily available for D2D applications to this day---interestingly, all of them are out-band.
Channels are arranged in families, either because they employ similar spectrum, or because they emanate from the same family of standards.
For each standards' family, several specifications were displayed to show the evolution of their characteristics, as specification efforts progressed.

\subsection{Out-band channels}\label{ssec:outband_channels}
\subsubsection{Free-Space Optical communication (FSO)}
Free-space optical communication has been used by humanity since its infancy---from smoke signals, to optical telegraphs, and the most recent visual light communication technologies.

\begin{table}
\caption{\label{tab:channels}Characteristics of the most widely available D2D communication channels}
\centering\small
\begin{tabular}{@{} p{3.3cm} p{2cm} p{1.7cm} p{2.6cm} p{2.4cm} l @{}}
\toprule 
Name & Specification date & Device\newline discovery\,$^*$ & Maximum\newline distance\,$^{**}$ & Maximum\newline data rate & Frequency \\
\midrule

\textbf{Optical} \\
Infrared (IrDA) & 1994~\cite{infrared_data_association_serial_2001} & Manual & Several meters & 16~Mb/s & 0.3-430~THz \\
\Screentocamera & --- & Manual & Around a meter & 23.6~kb/picture & 440-790~THz \\

\midrule
\textbf{Wi-Fi}  \\

802.11 (legacy) & 1997~\cite{ieee_802.11_1997} & Automatic & 100~m & 1-2~Mb/s & 2.4~GHz \\
802.11ax (Wi-Fi 6E) & 2021~\cite{ieee_802.11ax_2021} & Automatic & 100~m & 1.2~Gb/s & 1-7.125~GHz \\

802.11ay & 2021~\cite{ieee_802.11ay_2021} & Automatic & 10~m & 100~Gb/s & 60~GHz \\

\midrule
\textbf{Bluetooth} \\ 

Bluetooth Core v1 & 1999~\cite{bluetooth_v1_spec} & Automatic & 100~m & 1~Mb/s & 2.4~GHz \\
Bluetooth Core v5 & 2016~\cite{bluetooth_v5_spec} & Automatic & 100~m & 1-3~Mb/s~\cite{bluetooth_overview_2021} & 2.4~GHz\\

\midrule
\textbf{RFID} \\
NFC & 2004~\cite{iso_nfc} & Manual & 20~cm & 106-424~kb/s & 13.56~MHz \\

\midrule
\textbf{Utra-WideBand} \\
UWB & 2007~\cite{iso_uwb_standard_2007,iso_uwb_standard_2007b} & Automatic & 200~m & 27~Mb/s & 3.1-10.6~GHz \\

\bottomrule
\end{tabular}

\vspace{1ex}\raggedright\small
$^*$ Some channels require \emph{manual} human supervision to be used, whereas some others \emph{automatically} discover peers in the vicinity.

$^{**}$ The range of a radio signal depends on its frequency (which drives signal attenuation) and on the antenna's power and type (directional or omni-directional), not on its underlying standard~\cite{bluetooth_range_2021}.
The maximum distances presented here refer to typical user equipment.
\end{table}

\paragraph*{Infrared communication} 
Infrared refers to the electromagnetic spectrum between 300~GHz and 430~THz (visible red light), and was first `discovered' in 1800 by astronomer Herschel~\cite{rowan-robinson_night_2012}.
Since then, it has found countless applications, from night vision to astronomy to communication.
As a consumer product, infrared is routinely used for e.g. televisions' remote control.
Nintendo featured infrared sensors in most of their gaming consoles: from the 1998's GameBoy Color up to the 2017's Switch~\cite{nintendo_systems_list}.
It seems, however, that only a few games ever made use of the feature, highlighting a developers' preference for other wireless communication channels.
Nevertheless, to this day, several smartphone manufacturers continue to ship infrared `blasters' in their products,
Xiaomi notably provides a TV remote control application leveraging their smartphones' infrared sensors~\cite{xiaomi_inc_mi}.
Infrared communication requires that the emitter and receiver be in direct line of sight.
Depending on the source intensity, the maximum communication distance can span miles.
Most of infrared's commercial applications employ proprietary protocols, although the \emph{Infrared Data Association} (IrDA) provides a standard protocol since 1994~\cite{infrared_data_association_serial_2001}, advertising a data rate of up to 16~Mb/s.

\paragraph*{\Screentocamera}
Recent years have seen the advent of a one-way D2D communication channel---so ubiquitous it does not even have a name: shooting a QR code from a device's screen to another device through its camera.
The practice arises from QR codes, a well-settled technology that has already found worldwide adoption.
As a replacement to lower-capacity barcodes, QR codes are routinely used for product tracing, loyalty programs and the like.
It enables augmented reality applications, to display restaurants' menus or to locate objects in a 3D space.
Finally, in recent years, the \screentocamera channel has found massive public adoption, as it is pervasively used as an authentication technique for, e.g., travel tickets, Wi-Fi log-in and, most recently, Covid-19 contact tracing and access control~\cite{neil_campbell_qr_2021}.
According to its ISO specification~\cite{iso_qrcode_2015}, a QR code is capable of holding up to 23.6~kb of data---depending on the code's size, data type, and level of \emph{Error Correction Coding} (ECC).
Specifically, as a means to transfer information from a screen-enabled device to a camera-enabled one, it intuitively sounds impractical to use the \screentocamera D2D channel as a data streaming interface. 
Indeed, generating, capturing and interpreting a QR code is somehow cumbersome and time-intensive.
For the time being, let us only consider the channel as a means to exchange one-way \emph{pulses} of data.

\subsubsection{Wi-Fi}
In 1996, Chai-Keong Toh deposited a patent for routing ad~hoc mobile networks, published in 1999~\cite{toh_routing_1999}.
In 1997, the IEEE 802.11 Working Group published its first standard for wireless network communication in the 2.4~GHz frequency band~\cite{ieee_802.11_1997}, inside which the `ad~hoc network' mode was a first-class citizen.
Considered as `beta', this initial standard was superseded in 1999 by 802.11a~\cite{ieee_802.11a_1999} and 802.11b~\cite{ieee_802.11b_1999}, which found a massive adoption by manufacturers, and finally allowed `Wi-Fi' to become the ubiquitous wireless medium that we know.

`Ad~hoc networking' enables direct communication between end-devices, that is D2D communication.
Additionally, it supports multi-hop routing: two out-of-range devices can communicate through Wi-Fi ad~hoc by using an intermediary gateway device, effectively enabling \emph{Mobile Ad~hoc Networks} (MANETs)~\cite{corson_mobile_1999}---which is outside the scope of this article.
Lastly, Wi-Fi ad~hoc features the mesh topology, where any two devices connected to a Wi-Fi ad~hoc network can directly communicate with one another, without going through the gateway node, as long as they are in range.
It is notable that D2D communication was seriously considered by pioneers of Wi-Fi from its inception, and even beforehand.
Wireless network card manufacturers did implement the functionality, resulting in academic experiments on ad~hoc networks from 2000 onwards, notably by the aforementioned patent holder C.-K.~Toh~\cite{toh_experimenting_2000,toh_evaluating_2002}.
Because `Wi-Fi ad~hoc' exists since Wi-Fi's inception, it features the same capabilities as the most recent Wi-Fi standards' implementations: as of 2021, the information rate climbs up to 20~Gb/s at close range using the 60~GHz band (standard 802.11ay~\cite{ieee_802.11ay_2021}), while it maxes out at $9,608$ Mb/s on the traditional 4.2--6~GHz band (standard 802.11ax, also called Wi-Fi~6E~\cite{ieee_802.11ax_2021}).
Concurrently to IEEE 802.11's standardization efforts, the Wi-Fi Alliance periodically proposes standards enabling new applications and improved user experience with Wi-Fi.

\paragraph*{Wi-Fi Direct} 
`Wi-Fi Direct'~\cite{wi-fi_discover_direct} (formerly `Wi-Fi P2P') was introduced in 2010 by the Wi-Fi Alliance, in an initiative to make ad~hoc networks simpler.
This specification makes Wi-Fi operations closer to Bluetooth for use cases like device pairing, with the superior bandwidths of the Wi-Fi family.
Wi-Fi Direct restricts itself to single-hop communication in a star topology (all communications go through the Wi-Fi Direct gateway, coined the Group Owner).
By re-using Wi-Fi features from previous norms, only one of the Wi-Fi devices needs to be compliant with Wi-Fi Direct.

\paragraph*{Wi-Fi Aware}
`Wi-Fi Aware'~\cite{wi-fi_discover_aware} was introduced in 2015, mostly as a tool aggregating Wi-Fi and \emph{Bluetooth Low Energy }(BLE) to deliver energy-efficient advertisement \& discovery of services in proximity.
Users can read insightful data from `beaconing' devices (i.e. devices only emitting short pulses of information) without connecting to them, although the specification also allows switching to a connected mode arranged in a mesh topology (no group owner).
The Android kernel supports Wi-Fi Aware since 2017~\cite{google_wi-fi-aware}, hinting that the technology might have been considered by Google as a beaconing back-end. 
However, no other OS supports Wi-Fi Aware, effectively making it an Android-specific artefact.

\subsubsection{Bluetooth (BT)}\label{sssec:BT}
Bluetooth is another ubiquitous wireless communication channel utilizing the unlicensed 2.4~GHz radio band.
Following its first specification by the Bluetooth \emph{Special Interest Group} (SIG) in 1999~\cite{bluetooth_v1_spec}, the first Bluetooth-enabled consumer device was a hands-free headset, highlighting BT's core objective: to obsolete cables for connecting digital peripherals---a typical D2D use case.
At the time of writing, Bluetooth 5.4, published in February 2023, is the latest of 14 \emph{core} specifications of the protocol~\cite{bluetoothsigBluetoothCoreSpecification2023,bluetooth_overview_2021}.
BT standards are split into tens of \emph{profiles}~\cite{wikipedia_bt_profiles_2022}, segregated per usage scenario.
Profiles are not necessarily tied to a specific core specification.
Each BT-enabled device only needs to support a subset of the available profiles---e.g. `\emph{Advanced Audio Distribution Profile} (A2DP)' for a wireless headset---facilitating Bluetooth implementation for device manufacturers.
BT creates point-to-point connections, called \emph{pairs}, having a primary/secondary relationship.
Bluetooth connections support up to 1--3~Mb/s date rates.
A primary device can concurrently maintain connections with up to seven secondary nodes, resulting in a star-topology network coined a `piconet'~\cite{sairam_bluetooth_2002}.

\paragraph{Bluetooth Low Energy (BLE)} 
BLE is a protocol stack that allows BT to operate on small-size devices with very low power requirements (0.01--0.50~W).
It enabled the cheap manufacture of `beacons': small devices only used for broadcasting small bursts of information (they never create a full connection).
This class of small-scale devices opened the door to new use cases, such as indoor geolocation (where BLE beacons are scattered in an indoor location to allow Bluetooth-enabled devices to locate themselves).
BLE being a different protocol from `classical Bluetooth', they are incompatible. 
However, most appliances supporting the latter (e.g. smartphones) are `dual-mode': they also implement the BLE protocol.

Bluetooth and Wi-Fi may appear to be competing for the wireless consumer market:
Wi-Fi supports much higher data rates, while BT's profiles make the latter cheaper and more straightforward to implement on low-end IoT devices.
However, several proposals leverage both channels (e.g. Wi-Fi Aware), which hints that they might just as well be used in combination to improve the overall D2D connectivity.

\subsubsection{Radio-Frequency IDentification (RFID)} 
Operating on MHz bands with a data rate in the order of kb/s, RFID is a particularly cheap short-range wireless channel.
It has been commercially deployed for decades as a remote identification technology for, e.g., theft prevention, goods inventories \& access-control through badges~\cite{Use_of_RFID}.
This is thanks to RFID `tags', which are able to emit beacons using only the RF energy harvested from their environment, without a battery~\cite{Weis_rfid,Intel_rfid}.
In 2004, the ISO standardised NFC~\cite{iso_nfc}, a novel communication protocol based on RFID.
Most modern smartphones nowadays fully support NFC.
Operating at 13.56~MHz with a maximum throughput of 424~kb/s, NFC allows contactless communication between devices over a few centimetres.
This short range requires little power, and eschews interference issues.
Due to its low throughput, NFC is a bad fit for demanding data exchanges.
However, it is a very compelling channel for secure authentication, as emphasized by its broad adoption for contact-less payment and the newest Wi-Fi authentication schemes (see section~\ref{ssec:d2d_sec_features}).

\subsubsection{Ultra-WideBand (UWB)}
UWB is characterized by its wide frequency band (500~MHz).
It transmits several bits of information in one \emph{pulse}, over a range of frequencies.
This operating mode differentiates it from most other wireless channels, that exchange a bit per time step over a single frequency.
Despite being one of the first radio communication channels, UWB was only granted an authorization for commercial use by the FCC in 2002, on the unlicensed 3.1--10.6~GHz frequency band~\cite{pirch_introduction_2020,wireless_lan_professionals_ultra_2020}.
The ISO publishes UWB standards since 2007~\cite{iso_uwb_standard_2007,iso_uwb_standard_2007b}.
It is sponsored by the UWB Alliance and the FiRa Consortium since 2019.
IEEE's working group 802.15 on \emph{Wireless Personal Area Networks} (WPANs) is working towards a specification since 2020.
Deployment-wise, the iPhone 11 (2019) was the first consumer mobile device to support UWB~\cite{apple_support_ultra_2021}, thanks to Apple's U1 chip (see section~\ref{sssec:apple}).

As a D2D communication channel, UWB is advertised as a low-energy, high-throughput (27 Mb/s) channel supporting a range of several hundred meters.
It is also expected to become a precise geo-location and tracking technology~\cite{wireless_lan_professionals_ultra_2020}.

\subsubsection{Discarded channels}\label{sssec:atypical_channels}
According to our definition of D2D communication, ZigBee~\cite{ieee_802.15.4}, Li-Fi~\cite{islim_modulation_2016}, LoRa~\cite{augustin_study_2016} and SigFox~\cite{lavric_sigfox_2019} are outside the scope of this article:  they require dedicated hardware that is not widely available on consumer mobile devices.
Some of these protocols additionally require a third-party coordinator to exchange data between two peers.
Nevertheless, they are worth considering in broader D2D studies for several reasons:  Li-Fi eschews radio interferences by working on the visible light spectrum; the \emph{Low Power Wide Area Network} (LPWAN) properties of LoRa \& SigFox are particularly interesting for the IoT sector~\cite{mekki_overview_2018} and geo-location~\cite{aernouts_sigfox_2018}; and ZigBee---intended to be a simpler protocol than BT \& Wi-Fi---can be a good fit on low-end embedded devices (e.g. sensors).

\subsection{In-band channels}\label{ssec:inband_channels}
Cellular communication is regulated by the global standards organization 3GPP (for `3rd Generation Partnership Project').
The 3GPP began working on D2D communication in 2011, as part of their Release 12 (or Rel12)~\cite{3GPP_Rel12}, which was frozen in 2015.
The focus was put on public safety \& critical communication sectors (as a replacement for the legacy Land Mobile Radio channel or LMR), resulting in two 3GPP `Work Items': \emph{Proximity Services} (ProSe), and Group Communication.
The integration of emerging D2D use cases in further 3GPP releases continues as time passes. 
The latest frozen release, Rel15~\cite{3GPP_Rel15} notably took interest in \emph{Internet of Things} (IoT), \emph{Machine Type Communication} (MTC) and \emph{Vehicle-to-Everything} (V2x) use cases.
Although 3GPP protocols, such as \emph{LTE for Machines} (LTE-M) and \emph{NarrowBand-IoT} (NB-IoT), support direct communication between devices (without going through any cellular base station)~\cite{jameel_survey_2018,kar_critical_2020}, hardware support seems entirely absent from end-user appliances.

\subsection{Proprietary solutions}\label{ssec:proprietary}
Specialized devices typically feature ad-hoc wireless communication protocols, although generally not available on mainstream end-user equipment.
An iconic example is Apple, which is well-known for streamlining their users' experience across devices (e.g. iPhone, iMac, Apple TV...) through proprietary protocols.
Particularly interesting in the context of D2D communication are the AirPlay \& AirDrop services, due to their wide adoption and acknowledged performance (as observed in section~\ref{sec:d2d_survey}).
AirPlay~\cite{apple_airplay_2022}, dating back to 2004, only specialises in streaming audiovisual content from computers (including phones and tablets) to TVs and speakers.
AirDrop, on the other hand, is a general purpose file transfer solution. 
It was initially released in 2011, and can only be used between Apple devices.
Both services are backed by the undocumented, proprietary \emph{Apple Wireless Direct Link} (AWDL) protocol, and leverage standard close-range wireless channels: Wi-Fi \& Bluetooth, and UWB since iPhone 11~\cite{wireless_lan_professionals_ultra_2020}.
The \emph{Open Wireless Link} (OWL) project~\cite{owl:project} has been reverse-engineering AWDL since 2018~\cite{stute_one_2018}.
The OWL project's continued work include the publication of security flaws inside AWDL~\cite{stute_billion_2019} and an open implementation of AirDrop, coined OpenDrop~\cite{secure_mobile_networking_lab_opendrop_2022}.

Since 2018, Google also proposes a proprietary D2D solution for Android phones, entitled \emph{Google Nearby Connections} (GNC)~\cite{google_nearby}. 
As part of the Android Play Services, it is therefore installed in most Android phones.
Contrarily to AirDrop, GNC exposes APIs to application developers, allowing any Android app to establish general-purpose direct connections between phones (and even groups of phones).
Antonioli~\emph{et~al.}~\cite{antonioli_nearby_2019} reverse-engineered GNC in 2019, exposing severe security breaches in the same paper.
The paper shows that GNC operates similarly to AWDL.

Because they are undocumented and only leverage standard channels, AirDrop and GNC were left-out from table~\ref{tab:channels}.
However, these two proprietary channels still demonstrate that performant and reliable D2D communication is feasible, and mostly seems to demand a concerned OS support.

\subsection{Hardware}\label{ssec:hardware}

\subsubsection{General case}
Since 2007's iPhone 1~\cite{iPhone1}, smartphones have always supported at least cellular network, Wi-Fi, and Bluetooth communication.
At the time, mobile phones used to contain one chip per channel.
Nowadays however, most mobile phones contain a single circuit chip (a \emph{System on a Chip} or SoC) that integrates most of the features needed for a smartphone---i.e. CPU, GPU, volatile memory, network connectivity, camera\dots
One of the most widely deployed SoC as of 2023 is the SnapDragon series from Qualcomm that supports Bluetooth, Wi-Fi, cellular, and NFC communication. 
Another one is the Dimensity series by MediaTek, featuring the same connectivity support.
Since 2021, Google designs its own SoC, coined Tensor, which is shipped inside the Google Pixel series of smartphones, from version 6 onwards~\cite{nievaGooglePixelPhones2021}. 
Guessing from a system parameter to disable UWB on Pixel 6 Pro~\cite{liPixelProUWB2021}, the Tensor SoC supports UWB additionally to the above-mentioned channels.
Lastly, some SoC with no connectivity support also exist, such as the Kirin by HiSilicon. 
Smartphones equipped with the latter host additional hardware for wireless communication.

In the end, all smartphones support Wi-Fi and Bluetooth, the most ubiquitous out-band channels for D2D communication.
Hardware support for UWB is still nascent.
None of the current SoC for smartphones support in-band channels, such as LTE-M or NB-IoT (cf. sec.~\ref{ssec:inband_channels}), nor atypical channels like ZigBee (sec.~\ref{sssec:atypical_channels}).

\subsubsection{The Special case of Apple}\label{sssec:apple}
Apple devices are now fully designed by Apple: from the hardware, to the OS, to the system APIs.
Nevertheless, Apple's developments are generally closed-source, making it hard for the most curious of us to scrutinize their products' operations.
As presented in section~\ref{ssec:proprietary}, Apple's AirPlay and AirDrop D2D protocols are backed by the (AWDL) D2D protocol, which functions atop standard Wi-Fi chips according to the OWL project~\cite{owl:project,stute_one_2018}.
Since 2019, iPhones are packaged with the U1 network chip~\cite{u1_chip}, which additionally supports the \emph{Ultra-WideBand} (UWB) channel~\cite{barrett_biggest_2019}.

\section{Security models}\label{sec:security_models}

Compared to the Internet, the unique properties of D2D communication impose specific security challenges.
On one hand, D2D locality obviates most remote attack scenarios, which are daunting in the Internet paradigm.
On the other hand, the broadcast nature of D2D wireless channels and the need for periodic self-advertisement introduce new classes of attacks.
This section overviews the security model of D2D communication.
Several surveys propose a more in-depth study~\cite{haus_security_2017,wang_survey_2017}.
Section~\ref{sec:attack} introduces D2D attack scenarios, before enumerating desired security properties.
Section~\ref{ssec:d2d_sec_features} presents D2D-specific security features.
Finally, as an example of these features, we briefly describe Bluetooth and Wi-Fi current authentication schemes in Section~\ref{ssec:auth_wifi_bt}.

\subsection{Attack scenarios}\label{sec:attack}

\paragraph*{Passive attacks}
Due to the broadcast nature of wireless communication, an adversary \textit{in the vicinity of a user} may passively eavesdrop and/or analyse D2D traffic to learn sensitive information about other peers.
Unencrypted messages are directly accessible through \emph{eavesdropping}, while \emph{traffic analysis} can infer context information, even on encrypted traffic, such as the object of a transmission (text, file, stream\dots) or the service being used.
Location information can also be passively inferred, such as D2D users' whereabouts and travel patterns~\cite{primault_long_2019}.

\paragraph*{Active attacks}
An active adversary may perform a \emph{Denial of Service} (DoS) attack by \emph{jamming} the communication medium.
By \emph{forging} and/or \emph{manipulating} D2D messages, a malicious user in transmission range can further trick honest peers: they may access unauthorized services, impersonate the identity of other persons, or even steal information from their victims.

\subsection{Desired security \& privacy properties}
D2D security as a whole is driven by network security \& privacy requirements.
The two concepts are very interrelated, although they sometimes conflict.

\paragraph*{Network security}
The three usual tenets of network security are: 
\textbf{availability}, the fact of ensuring that the network is always accessible;
\textbf{integrity}, which protects users against the modification or falsification of exchanged information;
and \textbf{confidentiality}, ensuring that only authorized users can read a piece of information~\cite{shen_secure_2014}.
This latter is enabled by \emph{cryptographic authentication}, which provides for both \emph{identification} and \emph{encryption}~\cite{shiu_physical_2011}.
Encryption is of paramount importance in wireless, as any eavesdropper in range can read unencrypted traffic.

\paragraph*{Privacy}
Privacy does not hold any widely accepted definition, but can be defined as a personal decision to define which personal information one agrees to share with whom, how, and to what extent.
There is an inherent trade-off between privacy and utility, as the utility of a service generally increases as it is provided with more personal information on its user~\cite{haus_security_2017}.
In D2D communication, the major pillars of network privacy are:
\textbf{user anonymity}, or hiding from unauthorized peers any information that could be used to find one's identity;
\textbf{context privacy}, which prevents attackers from learning context information about a communication (e.g., location, service being used, query type);
and \textbf{unlinkability}, that aims at making several communication sessions by the same user indistinguishable from sessions by different users 
(one's communication history cannot be \emph{linked}).
As opposed to user anonymity---that takes interest in the messages' \emph{content}, unlinkability takes interest in the \emph{way} messages are sent~\cite{lindell_anonymous_2011}.

\subsection{D2D-specific security features}\label{ssec:d2d_sec_features}
\begin{figure}[t]
	\centering
	\includegraphics[width=0.7\columnwidth]{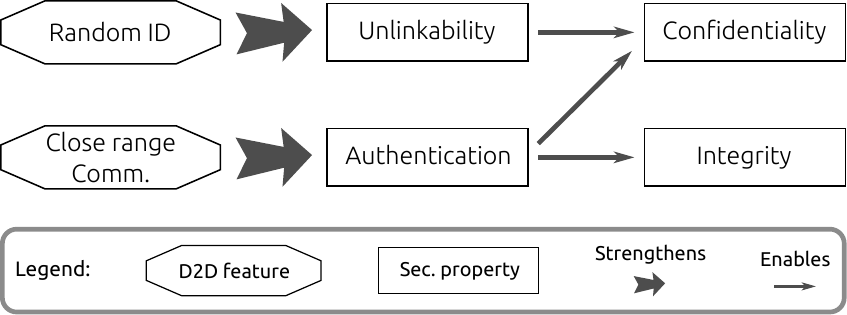}
	\caption{Some of D2D communication's features strengthen our desired security properties.}
	\label{fig:security_diagram}
\end{figure}

Figure~\ref{fig:security_diagram} depicts how D2D-specific features enable the aforementioned security properties.
We focus on network security properties, and not privacy ones, as the former knows more consensual definitions.

\paragraph*{Randomised host identifiers (Random ID)}
Internet communication stands atop the IP protocol of the third `network' OSI layer.
In the IP protocol, the IP address behaves like one's postal address on the net.
It is a mandatory piece of metadata to provide while communicating: without it, your addressee cannot answer back to you.
However, this piece of metadata jeopardises \emph{unlinkability}, because it identifies a user, and can thus easily be used for tracking individuals online~\cite{mishra_dont_2020}.
Hiding one's IP address remains possible through anonymity networks like Tor~\cite{dingledine_tor_2004}, but it is a cumbersome process, mostly undertaken by people in need for serious anonymity (e.g., journalists and whistleblowers).

On the other hand, we defined D2D communication as `single hop' communication, taking no interest in routing messages through networks: 
D2D communication stands below, in the first (`physical') and second (`data link') network layers.
Most often, data link is achieved through the MAC protocol.
Similarly to IP, a layer-2 address is required to identify communicating hosts, notably in the advertising (broadcast) phase.
Privacy-wise, this situation is bad, because such an address identifies a network card---thus its host device, and consequently its user.
However, \emph{there is no crucial need to broadcast one's \emph{real} layer-2 address to establish and maintain a D2D connection.}
Instead, this address can be randomly generated and periodically rotated by their host device to strengthen \emph{unlinkability}.
Efforts are indeed undertaken to hide this personal identifier by the most famous D2D channels:

\begin{itemize}
	\item In the Wi-Fi realm, the effort of randomising the MAC address has initially been made by OS manufacturers, such as Linux Tails~\cite{tails_mac_2021}.
	Although, IEEE~802.11 has been pushing MAC randomization since 2018 (as part of amendment 802.11aq), and created a task group entitled `Randomized and Changing MAC Addresses' in 2021, specifically interested in anonymising MAC addresses without loss of utility for related services~\cite{ieee_80211_random_mac_2022}.
	Indeed, several network operations traditionally depend on having fixed MAC addresses, such as MAC filtering or DHCP reservation;
	\item The Bluetooth 4.0 specification introduced the `smart privacy' feature: it notably replaces one's broadcast MAC address with a random one, derived from the real Bluetooth card MAC address~\cite{bluetooth_privacy_2015}.
	The only way to retrieve the real MAC from the random one, is using the broadcaster's \emph{Identity Resolution Key} (IRK).
	The IRK is only shared with remotes after pairing, meaning that only trusted peers are able to de-anonymise one's broadcast MAC address.
\end{itemize}

Having unlinkable communication sessions benefits \emph{confidentiality}, because it hides real network identifiers from unauthorized parties.
Furthermore, such randomized \& evolving addresses preclude longitudinal location tracking, and thus also improve \emph{context privacy}.

\paragraph*{Close range communication}
Digital authentication is the process of verifying one's identity, to ascertain that they are authorised to access specific pieces of information.
Authentication revolves around the exchange and validation of cryptographic keys, that are then used to encrypt and sign future messages.
To that regard, authentication enables both the \emph{confidentiality} and \emph{integrity} security properties.

Three factors are generally used to establish one's digital identity:
\textbf{knowledge}, where the system verifies that the user \emph{knows} a private information (password, PIN codes...);
\textbf{possession}, where the system verifies that the user \emph{owns} a physical object (computing device, token...);
and lastly the \textbf{biometric factor}, where the system verifies something that the user \emph{is} (fingerprints, face, eye scan...).
Using Wi-Fi or Bluetooth, we are used to exchanging passwords, which constitute \emph{knowledge factors}. 
Yet, authentication can be further facilitated by using physical proximity to assert a \emph{possession factor}.
\emph{Close range communication} (e.g., NFC or QR code scanning) is a convincing way to ensure that a communication is initiated between two legitimate users, as both assert that they are the ones \emph{owning} the communicating devices.
Furthermore, close range communication can be viewed as a bootstrap side-channel to exchange encryption keys, which eschews the need for e.g. Diffie-Hellman key exchanges (prone to man-in-the-middle attacks), and establishes a second authentication factor: \emph{knowledge}.

Note that authentication does not necessarily imply the disclosure of one's \emph{full} identity.
Through cryptography, and with the help of a source of trust (such as a certificate authority), it becomes possible to ensure that one is authorized to access a service, without disclosing any other information. 
The process is called \emph{anonymous authentication}~\cite{lindell_anonymous_2011,feng_anonymous_2020}, but cannot be easily carried out without the help of a third-party (outside the D2D communication scope).

\subsection{Authentication in Bluetooth and Wi-Fi}\label{ssec:auth_wifi_bt}
Since their inception, both channels allow for open or password-protected authentication of peers.
While Wi-Fi networks hold long-term passwords, in some cases Bluetooth employs one-time PIN codes (which only need to be matched on both pairing devices).
In Bluetooth, the initial authentication (pairing) is usually manual.
Because devices exchange encryption keys (such as the aforementioned IRK) while pairing, connecting already paired devices is automatic and much simpler: the devices recognize and connect to each other, without the need for human intervention.
In 2007, the Wi-Fi Alliance proposed \emph{Wi-Fi Protected Setup} (WPS) to ease network onboarding, notably featuring a PIN code scheme similar to Bluetooth's. 
Sadly, WPS' PIN code mode contained a vulnerability, allowing an attacker to brute-force the PIN code under certain conditions~\cite{viehbock_brute_2011}.

\emph{Wi-Fi Easy Connect} (WEC)~\cite{wi-fi_discover_easy_connect} was introduced in 2018 to further ease network authentication of interface-less IoT devices.
WEC connects such a device (called `enrollee') to a wireless network by using a third-party `configurator' (smartphone or else).
By scanning a QR code or NFC tag on the enrollee, the configurator initiates a secure connection with it to exchange information about the Wi-Fi access point and the connecting device.
Through the configurator, the enrollee then gets clearance, and can finally connect to the network by itself.
This process constitutes an example of close range communication for authentication, as discussed above.

\section{Conclusion}
\label{sec:conclusion}

This survey article took interest in the adoption state of D2D communication for end-users. 
We motivated the importance of this communication paradigm in terms of environmental sustainability, privacy and disparities in access to broadband Internet.
An online poll showed that usage scenarios where D2D communication would be beneficial exist in the wild. 
The state of the practice in communication channels was then reviewed, as well as the paradigm's security models. 
The path seems clear: device-to-device is needed and timely, many solutions exist, with interesting security features. 
So why such a low adoption rate? 

\begin{figure}[t]
    \centering
    \includegraphics[width=0.6\columnwidth]{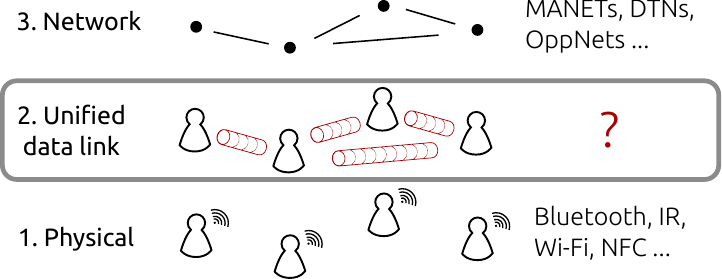}
    \caption{The literature lacks a standard abstraction to unify D2D channels in the link layer.}
    \label{fig:unified_l2}
\end{figure}

A major technical hurdle remains---D2D communication's underwhelming support in operating systems.
All existing APIs are either channel-specific and low-level or proprietary,
Apple has never allowed Bluetooth file transfer,
Android's Wi-Fi ad-hoc APIs are impractical (e.g. randomized hotspot SSID and password), 
etc.
The variety of Bluetooth's hardware \& software implementations is often pointed at to explain its underwhelming performances on certain device pairs.
Further study should be undertaken to quantify the extent of these alleged compatibility issues.
As illustrated in figure~\ref{fig:unified_l2}, 
the authors believe that a higher-level cross-platform API encompassing several communication channels would ease development.
Similarly to the IP stack, such an API would abstract the physical layer below more practical link-oriented controls.
Indeed, all D2D channels follow the same communication process: discovery (beaconing \& scanning), optionally followed by an authorization and a connected phase.
Going a step further, a unified network layer would be beneficial for transparent multi-hop D2D routing.
Two decades of research on infrastructure-less networks have scrutinized the topic, e.g. \emph{Mobile Ad~hoc Networks} (MANETs), \emph{Delay-Tolerant Networks} (DTNs), and \emph{Opportunistic Networks} (OppNets)~\cite{corson_mobile_1999,cc_survey_2016,trifunovic_decade_2017}.
We hope that this article motivates the adoption of more local communication processes, 
and provide a solid ground to continue advancing toward this end.

\printbibliography

\end{document}